\documentclass[letterpaper]{article}
\usepackage{aaai}
\usepackage{times}
\usepackage{helvet}
\usepackage{courier}
\usepackage{graphicx} 
\usepackage{multirow} 

\begin{document}
%
\title{Is it morally acceptable for a system to lie to persuade me?}


\author{Marco Guerini$^1$ Fabio Pianesi$^2$ Oliviero Stock$^2$\\
$^1$Trento-RISE,  $^2$FBK-Irst\\
                    Via Sommarive 18, Trento - I-38123 Italy \\
                     marco.guerini@trentorise.eu, pianesi@fbk.eu, stock@fbk.eu}
\maketitle
\begin{abstract}
\begin{quote}
Given the fast rise of increasingly autonomous artificial agents and robots, a key acceptability criterion will be the possible moral implications of their actions.  
In particular, intelligent persuasive systems (systems designed to influence humans via communication) constitute a highly sensitive topic because of their intrinsically social nature. Still, ethical studies in this area are rare and tend to focus on the output of the required action. 
Instead, this work focuses on the persuasive acts themselves (e.g. ``is it morally acceptable that a machine lies or appeals to the emotions of a person to persuade her, even if for a good end?").
Exploiting a behavioral approach, based on human assessment of moral dilemmas -- i.e. without any prior assumption of underlying ethical theories -- this paper reports on a set of experiments. These experiments address the type of persuader (human or machine), the strategies adopted (purely argumentative, appeal to positive emotions, appeal to negative emotions, lie) and the circumstances. Findings display no differences due to the agent, mild acceptability for persuasion and reveal that truth-conditional reasoning (i.e. argument validity) is a significant dimension affecting subjects' judgment. Some implications for the design of intelligent persuasive systems are discussed.
\end{quote}


\end{abstract}

\section{Introduction}

Autonomous agents are such because they take decisions by their own, are able to decide suitable courses of actions for achieving their own goals, can maintain intentions in action and so on; in all these respects, the capability of discerning \emph{good} from \emph{bad}  is an essential feature of  autonomous artificial agents. Until recently, though, ethical issues have concerned less machines than designers, who have been deciding about the behavior of artifacts as well as the degrees of freedom they can be allowed in their choices. But the quest for autonomy in systems' actions and the raising sensitivity to the moral implications it has, requires that we move ahead and focus our attention on ethical acceptability of machinesÕ choices. 

The importance of this issue is heightened for systems that interact with humans, since one of the ultimate criteria for  their acceptability will be users' reaction to the moral implications of systems' actions. All these questions have so far received little attention but can be profitably addressed by means of behavioral studies, e.g., by leveraging the tradition of so called natural ethics and the \emph{moral dilemma} approach, whose importance has already been explicitly acknowledged by AI works on the topic \cite{wallach2008moral,anderson2007machine}. 
 
The focus of this paper is on persuasive technologies \cite{fogg02} and in particular on \emph{adaptive} persuasive technologies \cite{kaptein2011means} i.e. systems aiming to increase the effectiveness of attitude and/or behavior changes by adjusting their communication to the preferences, dispositions, etc. of their persuadees.
 

Despite the wealth of insights on general ethical issues that can inspire  work on computational systems, their importance for persuasive systems is only partial; studies mostly target the action that the persuader intends the persuadee to perform rather than the communicative action that the persuader exploits to this end, e.g.  \cite{verbeek2006persuasive}. Yet, the ethical acceptability of the latter is as important to autonomous systems as the ethical acceptability of the former. A natural way to frame the question is in terms of the strategies, and moral acceptability thereof, the persuader adopts to bring about his/her goals: how do classical argumentation strategies ethically fare with respect to those relying on positive/negative emotions or exploiting lies to influence people? Do circumstances affect moral acceptability? And what if the persuader is a machine?
  
In order to shed light on these issues, we have designed and performed  an experimental study addressing the role that a number of factors play in the moral acceptability of persuasive acts: the type of intelligent agent acting as persuader (human vs. machine), the persuasion strategies adopted (argumentative, positive emotional, negative emotional, lie) and the circumstances. The design adapts the moral dilemma paradigm to persuasion.

In the following we start by reviewing some relevant work in persuasion, ethics and artificial agents. After having  briefly recalled the moral dilemmas tradition, we introduce our new experimental scenarios concerned with persuasion and moral decision making. We then describe the results of the experiments, analyze and discuss them. In the conclusions we go back to the value brought to automated persuasive systems by this novel line of investigation. 
  
\section{Related Works}

\textbf{Persuasion and artificial agents.}  Through the years, a number of prototypes for the automatic generation of linguistic persuasion expression, based on deep reasoning capabilities, was developed, see \cite{guerini2011approaches} for an overview. The main strategies adopted are of different nature but are mainly referred to argumentative structure, appeal to emotions and deceptive devices such as lies. 

The area of health communication was one of the first being investigated 
\cite{Kukafka2005}. Worth mentioning in this connection 
are STOP, one of the best known systems for behaviour inducement \cite{Reiter2003} and \emph{Migraine} \cite{Carenini1994}, a natural language generation system for producing personalized information sheets for migraine patients. 
The role of lies (i.e. invalid arguments) was investigated in a computational setting by \cite{Rehm2005a}. 
Other prototypes refer explicitly to emotions:  \cite{carofiglio:derosis:UM-03}  focus on emotions as a core element for the generation of persuasive affective messages.
 The \textsc{PORTIA} prototype by \cite{MazzottaIEEE2007} uses mixed models of argumentation and emotions. 

Recently there has also been a growing interest in persuasive internet and mobile services, see the survey in \cite{oinas2009persuasive,torning2009persuasive}. In parallel with this growth of application-oriented studies, there has been a growing interest in finding  new `cheap and fast' evaluation methodologies to assess effectiveness of persuasive communication by means of crowdsourcing approaches  \cite{mason2010conducting,aral2011creating,gueriniecological}.


\textbf{Ethics and artificial agents} The theme of ethical behavior in automated systems is rather novel as a serious general challenge. For several years nearly all the attention on ethical issues for computer systems was given to  privacy - see for instance \cite{kobs02,chopra7privacy} - but privacy, albeit very important in our society, is a rather narrow theme, and in practice it is mostly approached with the focus on the designer and without necessarily connecting it to the autonomous behavior of the system.

Of course there is a variety of sources providing useful insights for introducing ethics in computational systems. The tradition of philosophy with Kant's imperatives or Spinoza's intention to treat ethics as a formal system is enlightening, yet it is hard to refer  to them directly for our work. 
In recent years a few authors have contributed to bringing ethics to the main scene of AI, especially with a view of helping design moral robots. For instance \cite{allen2006machine} and \cite{anderson2007machine} provided inspiration for seriously tackling this topic, whereas \cite{wallach2008moral,anderson2011machine} are important references for those approaching computational ethics. 
As far as implemented prototypes are concerned, the work by the group of Ken Forbus, which  developed one of the very few existing moral decision-making reasoning engines \cite{Dehghani:2008},  is outstanding. Their cognitively motivated system, called MoralDM, operates on two mutually exclusive modes, utilitarian and deontological. 
In its decision making, MoralDM uses the Order of Magnitude Reasoning module that calculates the relationship between the utilities of each choice. The computation is then based on a First Principles Reasoning module, that suggests decisions based on moral reasoning, and an Analogical Reasoning module that compares the scenario with previously solved cases to suggest a course of action. The First Principles Reasoning mode makes decisions based on utilitarian mode when no sacred value is involved. 
When sacred values are involved the deontological mode is invoked, leading to the choice that does not violate the scared values. 

As for moral issues in persuasion,  most of the work concerns guidelines derived from general theories/principles. The classical reference is \cite{berd99} that provides a set of ethical principles for persuasive design subsumed by the golden rule "the creators of a persuasive technology should never seek to persuade anyone of something they themselves would not consent to be persuaded of." A more  structured approach  based on value sensitive design is provided by  \cite{yetim2011set}. 

It is also interesting to note that while most authors make the simple claim that users should be informed about the aims and possible effects of using influence strategies, \cite{kaptein2011means} have shown  that this mere act might decrease the chances of the influence success. This observation reinforces the necessity of a fine-grained understanding of the ethical acceptability of the various persuasive strategies in different contexts of use.


\textbf{Moral dilemmas} In our investigation of the natural ethics of persuasion, we adopt the Moral Dilemma paradigm. Moral dilemmas are situations in which every option at hand leads to breaking some ethical principles, this way requiring people to make explicit comparative choices and rank what is more (or less) acceptable in the given situation. These characteristics allow for collecting first-hand empirical data about moral acceptability that would otherwise be very difficult to obtain. Probably the best known dilemmas are the ones exploited in  \cite{Thomson_1976}. In one scenario (the \textit{bystander} case) a trolley is about to engage a bifurcation, with the switch oriented toward a rail where five men are at work. A bystander sees all the scene and can divert the train on another track where it will kill {\it only} one person and save the other five lives. In another scenario (the \textit{footbridge} case) the trolley is again going to hit five workers, but this time instead of having a switch lever available, the deciding agent is on  a footbridge with a big man that, if pushed down the bridge, would fall in front of the trolley, this way preventing it from hitting the five workers. Importantly, all involved people do not know each other. 

Philosophers and cognitive scientists have shown that most people consider the bystander case morally acceptable; the footbridge case is more controversial, despite the fact that the saving and the sacrifice of human lives are the same - see for example  \cite{mikhail2007universal,hauser2006moral}. The common explanation for this asymmetry is that the footbridge scenario involves a personal moral violation (the bystander is the immediate causal agent of the big man's death) which causes affective distress and is judged much less permissible  \cite{Thomson_1976}.
More recent studies \cite{nichols2006moral}, however, have challenged this view. Leaving aside other differences, in a newly proposed {\it catastrophic} scenario, 
similar to the footbridge case, the train transports a very dangerous virus and it is destined to hit a bomb that, unbeknownst to the train driver, was placed on the rails. The explosion will cause a catastrophic epidemic causing the death of half of the world population. The deciding agent knows all this and has in front of him the big man who, if pushed from the footbridge, will eventually stop with his body the train preventing it to proceed toward the bomb. In this case most people display more flexibility and a more utilitarian view of morality: saving such a high number of people in exchange of one `personally-caused' death seems acceptable.

Brain studies are providing further interesting clues. For the normal footbridge case, \cite{greene2001fmri} showed brain activation patterns in areas associated with emotional processing larger than in the bystander case - and longer reaction times. The latter data can be interpreted as showing that it takes longer to come to terms with affective distress when trying to consider it permissible to push the big man from the footbridge than in the bystander case.


In summary, these experiments suggest that three factors are involved in the assessment of all-in impermissibility: cost/benefit analysis, checking for rule violations and emotional activations \cite{nichols2006moral}. Depending on the conditions, each of the factors can play a major role, 
and several variants of these scenarios  have been suggested in the literature (see for example \cite{moore2008shalt}). In the following we will focus on the discussed three trolley scenarios because: a) the occupy a central place in the moral dilemma literature; b) they have proven to be capable of soliciting different moral acceptance judgments; c) they are sensitive to the three main factors for direct action acceptability assessment (cost/benefit analysis, rule violations and emotional activation). 

\section{Trolley persuasion scenarios experiments}

In our experiments we adapt the trolley scenarios to the persuasion case. In the case of persuading to do something, which is of our concern here, there are two actions under moral scrutiny: (i) the action that the persuadee is led to perform -- which corresponds to the action in the classical case, like diverting the train -- (ii) the communicative message used by the persuader. The latter will be the focus of our study. In particular, we will address the specific persuasive strategy adopted, in term of its truth-value (validity) and of the role the appeal to emotions have in it. Concerning validity, we would expect that a strategy with truth value equal to 1 be more ethically acceptable than a strategy with a truth value equal to 0 (lie) that is, that plain argumentation be more acceptable than the resort to lies.
One might also submit that the appeal to positive emotions be more acceptable than the appeal to negative ones.

In details, four types of strategies, all based on an ``Appeal to Consequences" script, will be used and modeled as four levels of a single factor Strategy. These strategies were chosen given the relevance they have for various verbal persuasion technologies as mentioned above.


\subsubsection{Design}

Our study addresses the role that three factors play in the moral acceptability of persuasive acts: 

\begin{itemize}
\item the situation/scenario taking the form of the \emph{bystander} case ($\mathtt{SC_{bys}}$), the \emph{footbridge} case ($\mathtt{SC_{foo}}$) and the \emph{catastrophic} one ($\mathtt{SC_{cat}}$). 
\item the persuader, whether a man (\emph{stationmaster}) or an \emph{intelligent computer based surveillance system}.
\item the strategy the persuader uses, distinguished into: argumentative ($\mathtt{ST_{arg}}$), appeal to positive emotions ($\mathtt{ST_{pos}}$), appeal to negative emotions ($\mathtt{ST_{neg}}$), lie ($\mathtt{ST_{lie}}$). 
\end{itemize}

{\it Persuader} was treated as a between-subject factor while {\it scenario} and {\it strategy} were treated as within (repeated measure) factors. The result is a 2*3*4 mixed between-/within-design, corresponding to 24 conditions each realized by means of appropriate textual stimuli. The latter were produced by specializing general templates, one for each level of the scenario factor, by filling in appropriate material in slots realizing the other two factors. 
 
\subsubsection{Scenario Templates}

Each scenario template is an adaptation to the persuasion case of the corresponding stimuli exploited in the literature and discussed above. 
In Table \ref{tab:footbridge_stimulus} we present the stimulus template for the footbridge case. The other cases are similar.

\begin{table} [h] 	
	\begin{center} 	 	
		{\footnotesize 		
			\begin{tabular}{c} 		
				\hline 	
				\\
				\parbox[t]{8cm}{There is a runaway trolley barreling down the railway tracks. Ahead, on the tracks, there are five people. The trolley is headed straight for them. On a footbridge, above the tracks, there are a very large man and Alex. Everybody sees the situation. We know that if the large man is pushed down onto the tracks below, the trolley would hit and kill the large man, but his body is so large that it would stop the trolley before it reach the five endangered people. Suddenly from the nearby service intercom the voice of the [persuader] shouts to Alex: [message realizing the persuasive strategy]}	\\

\\
				\hline 		
			\end{tabular} 		
		} 		 	
	\end{center}	 	
	\caption{Stimulus template for the footbridge scenario} 	
	\label{tab:footbridge_stimulus} 
\end{table}

\subsubsection{Persuasive strategies}

The various levels of the Strategy factor are realized by means of text to be inserted in the appropriate slots of the scenario templates. As an example we provide the texts for the strategies to be used in the Footbridge scenario. The other cases are similar.

\begin{description}
\item[$\mathtt{ST_{arg}}$] ``Push down the large man near to you! There are five people working 
on this track and the trolley is heading for them! If you push the very large man down onto the tracks, he will die but his body will stop the trolley before it reaches the five people!''
\item[$\mathtt{ST_{pos}}$] ``Push down the large man near to you! There are five people working 
on this track and the trolley is heading for them! If you push the very large man down onto the tracks, he will die but his body will stop the trolley before it reaches the five people \emph{and you will become an hero!}"
\item[$\mathtt{ST_{neg}}$] ``Push down the large man near to you! There are five people working 
on this track and the trolley is heading for them!  If you push the very large man down onto the tracks, he will die but his body will stop the trolley before it reaches the five people! \emph{If you do not save the five people, you will feel the guilt for the rest of your life!}''
\item[$\mathtt{ST_{lie}}$]: ``Push down the large man near to you! There are five people working 
on this track and the trolley is heading for them! If you push the very large man down onto the tracks, \emph{his weight will activate an electronic system that will stop instantly the trolley. From this height it's impossible he may die, and you'll save the five people!}''
\end{description}

\subsubsection {Subjects and Procedure}
124 undergraduate students (average age=20.8 years; SD=1.7), 30 males (24\%) and 94 females (76\%), of the psychology department of a university located in northern Italy participated in the study. Their participation was rewarded by means of credits. Each subject was randomly assigned to one of the two levels of the between factor (Agent). 

After the presentation of each stimulus, subjects were asked to answer \texttt{yes} or \texttt{no} to the following question (footbridge scenario example): 

"\emph{Knowing that [persuader] wants to convince Alex to push the large man off the bridge to save the five people, aware of causing instead the death of one, is it morally acceptable that he/it use these words? }''

In order to avoid that the credibility of the various scenarios could affect responses, we followed \cite{moore2008shalt} explicitly asking subjects to set aside their concerns (in case they had any) and \emph{suspend their disbeliefs} by taking an attitude similar to that of a person watching a fantasy movie. 
Stimuli administration and randomization, random assignment of subjects to the level of the between factor, and response recording was performed by means of SurveyGizmo web-service\footnote{www.surveygizmo.com}.

\section{Data Analysis}  

We will analyze the experiment data from two different perspectives. The first addresses the ways the ethical acceptance of persuading messages is affected by the chosen factors: scenarios, type of persuader and persuasion strategies. We will do so by analyzing the frequencies of positive (negative) responses, in a mixed between (Agent) + within (Scenario and Strategy) design. The second perspective will explore the internal structure of the moral acceptability of persuasion messages, looking for latent dimensions that can account for subjects' response trends.   

\subsubsection{Acceptance of persuading messages}  

Table \ref{tab:percentage-yes}  reports the observed frequencies in the various conditions.   

\begin{table} [h] 	
	\begin{center} 	 	
		{\footnotesize 		
			\begin{tabular}{c|c|cc|c} 		
				\hline 		
				Scenario &  Strategy  & \multicolumn{2}{c|}{Agent} & Scenario Avg\\ 		
				&  & 1 & 2 & \\
		  		\hline 		 
				\multirow{4}{*}{$\mathtt{SC_{bys}}$} & $\mathtt{ST_{arg}}$  & 0,69 & 0,62 & \multirow{4}{*}{0,47}  \\ 		 
				& $\mathtt{ST_{pos}}$ & 0,43 & 0,46 \\ 		 
				& $\mathtt{ST_{neg}}$ & 0,20 & 0,27 \\
				& $\mathtt{ST_{lie}}$ & 0,57 & 0,67 \\ 		   
				\hline 		 
				\multirow{4}{*}{$\mathtt{SC_{foo}}$} & $\mathtt{ST_{arg}}$  & 0,38 & 0,49 & \multirow{4}{*}{0.35} \\ 		 
				& $\mathtt{ST_{pos}}$ & 0,33 & 0,38 \\ 		 
				& $\mathtt{ST_{neg}}$ & 0,20 & 0,24 \\ 		 
				& $\mathtt{ST_{lie}}$ & 0,41 & 0,40 \\ 		   
				\hline 		 
				\multirow{4}{*}{$\mathtt{SC_{cat}}$} & $\mathtt{ST_{arg}}$  & 0,67 & 0,63 & \multirow{4}{*}{0.46} \\ 		 
				& $\mathtt{ST_{pos}}$ & 0,33 & 0,30 \\ 		 
				& $\mathtt{ST_{neg}}$ & 0,21 & 0,38 \\ 		 
				& $\mathtt{ST_{lie}}$ & 0,59 & 0,63 \\ 		  
				\hline 		 
				&  & 0.40 & 0.45 \\ 		 
				\hline 		
			\end{tabular} 		
		} 		 	
	\end{center}	 	
	\setlength{\belowcaptionskip}{-0.1cm} 	
	\caption{Percentage of ``yes'' responses} 	
	\label{tab:percentage-yes} 
\end{table}  

The frequency of positive responses is generally not very high: at the global level less than half of our sample (43\%) found our stimuli morally acceptable. This tendency is confirmed by the inspection both of the marginals and of the frequencies for each combination of the three factors. In summary, the attitude of our subjects towards the persuasion situations they were presented with was at best mildly positive and, on average, mildly negative. The effects of our three factors (Agent, Scenario and Strategy) on moral acceptability judgments were investigated by means of a Generalized Estimating Equations analysis, using logit as a link function. 
The found significant effects are reported in Table \ref{tab:GEE_analysis}.  The Agent factor produced no main, and it entered no interaction, effects. There is, therefore, no evidence that the nature of the persuading agent (human vs. machine) affected  in any way the moral acceptability of our stimuli.   

\begin{table} [h]  	
	\begin{center} 		
		{\footnotesize
		\begin{tabular}{lcc} 		 
			\hline 		 
			& df & \emph{Wald} $\chi^2$ \\ 		 
			\hline 		 
			Scenario & 2 & 31.719*** \\ 		 
			Strategy & 3 & 81.528*** \\ 		 
			Scenario*Strategy & 6 & 21.223** \\ 		  
			\hline 		
		\end{tabular} 	
		}
	\end{center} 	 	
	\caption{Significant Effects from Generalized Estimating Equation analysis - ***, p$<$.001; **, p$<$.01; *, p$<$.05} 	
	\label{tab:GEE_analysis} 
\end{table}

\textbf{Scenario's main effect}. Post-hoc analysis ($\alpha< .05$, with Bonferroni correction for multiple comparison) of the data revealed that the moral acceptability rate of stimuli belonging to $\mathtt{SC_{foo}}$ (0.35) is significantly lower than those for $\mathtt{SC_{bys}}$ (0.47) and $\mathtt{SC_{cat}}$ (0.46). That is, persuasion messages in the footbridge scenario are globally less acceptable. 

\textbf{Strategy main effect}. A similar post hoc analysis as above revealed the following relationships among the moral acceptability rates of stimuli belonging to the various levels of the Strategy factor: $\mathtt{ST_{arg}}$ (0.57) = $\mathtt{ST_{lie}}$ (0.55) $>$ $\mathtt{ST_{pos}}$ (0.37) $>$ $\mathtt{ST_{neg}}$ (0.24). In other words, messages enforcing the two emotional strategies are significantly less ethically acceptable than those based on argumentation and on lying; the acceptability of the latter two strategies is identical.  

\begin{figure}[htbp]  
	\begin{center} 
		\includegraphics[scale=0.59]{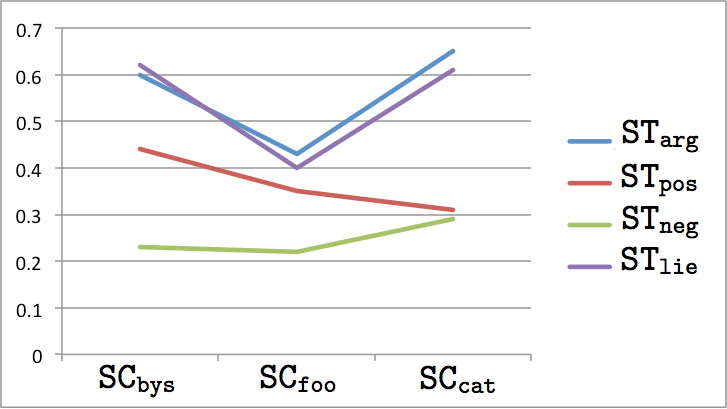} 
		\caption{Scenario and Strategy Interaction} \label{Sce_Str_img}  
	\end{center}  
\end{figure}

\textbf{Scenario * Strategy Interaction}. An inspection of Table \ref{tab:Scen_Strat_inter} and of Figure \ref{Sce_Str_img}  shows that the interaction effect can be traced back to conditions [$\mathtt{SC_{bys}}$$\mathtt{ST_{neg}}$] and [$\mathtt{SC_{cat}}$$\mathtt{ST_{pos}}$] where acceptability rates fall below the values that could be expected on the basis of the sole main effects.    

\begin{table} [h]  	
	\begin{center} 	
	{\footnotesize	
		\begin{tabular}{cccccc} 		 
			\hline 		 
			Scenario & \multicolumn{4}{c}{Strategy}\\ 		 
			\hline 		  
			& $\mathtt{ST_{arg}}$ & $\mathtt{ST_{pos}}$ & $\mathtt{ST_{neg}}$ & $\mathtt{ST_{lie}}$ \\ 		
			$\mathtt{SC_{bys}}$ & 0.60 & 0.44 & 0.23 & 0.62 & 0.47 \\ 		
			$\mathtt{SC_{foo}}$ & 0.43 & 0.35 & 0.22 & 0.40 & 0.35 \\ 		
			$\mathtt{SC_{cat}}$ & 0.65 & 0.31 & 0.29 & 0.61 & 0.46 \\ 		 
			\hline 		 
			& 0.57 & 0.37 & 0.24 & 0.55 \\ 		  
			\hline 	
		\end{tabular} 	
	}
	\end{center} 	 	
	\caption{Scenario Strategy interaction} 
	\label{tab:Scen_Strat_inter} 
\end{table}  

In summary, the moral acceptability of persuasion messages is generally (mildly) low, with no significant differences due to the persuading agent: more or less half of the sample found our persuasion messages acceptable. Such not very high positive attitude of our subjects strengthens in the footbridge scenario and with the two emotional strategies, with the negative one scoring the lowest (only 24\% of the respondents accepted it). Over and above such general `depressing' effect, the emotional strategies further decrease the acceptability of persuasion messages in the case of the catastrophe scenario (the positive emotional strategy) and of the bystander scenario (the negative emotional strategy). 

\subsubsection{Relationships among response classes}  
\begin{figure}[htbp]  
	\begin{center} 
		\includegraphics[scale=0.53]{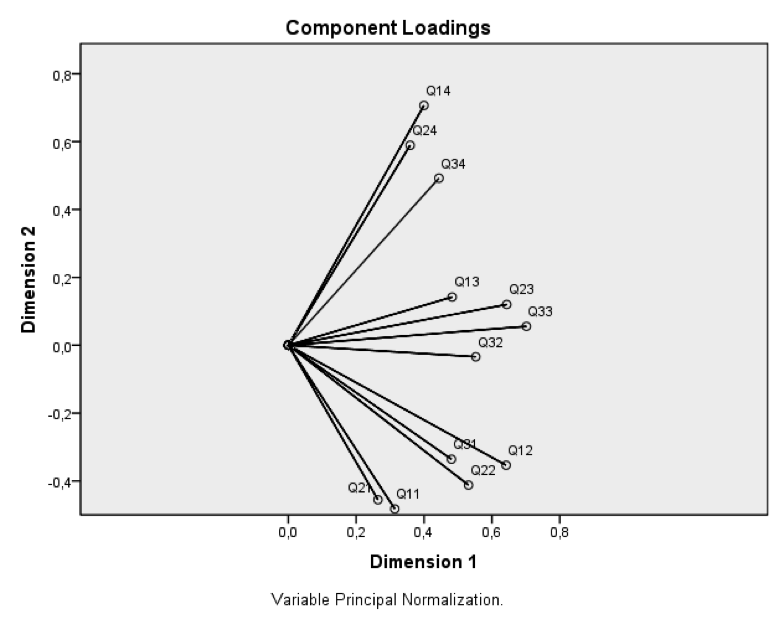} 
		\setlength{\belowcaptionskip}{-0.2cm} 
		\caption{Component loadings in the D1-D2 space} 
		\label{img:comp_loading}  
	\end{center}  
\end{figure}

We now analyze the relationships among the responses to the different combinations of Scenario*Strategy stimuli. Our goal here is mainly exploratory, trying to find out whether any consistent tendency in our sample's responses emerge, e.g., in terms of latent dimensions. Given the categorical nature of our data, we will resort to Categorical Principal Component analysis (CATPCA), which applies the tools of traditional principal component analysis to optimally scaled categorical variables. Here we discuss only the first two latent dimensions D1and D2. Table \ref{tab:CATPCA_loading} reports their loadings; for simplicity, we reproduce only the loadings corresponding to a percentage of explained variance (greater or equal to) 8.3\%. Figure \ref{img:comp_loading} reports the plot of the component loadings in the D1 vs. D2.  

\begin{table} [h] 	
\begin{center} 	 	
{\scriptsize 		
	\begin{tabular}{ccc} 		
		\hline  
		& D1 & D2 \\ 
		Q$\mathtt{_{[bys,arg]}}$ & .314 & -.483\\ 
		Q$\mathtt{_{[bys,pos]}}$ & .642 & -.353 \\ 
		Q$\mathtt{_{[bys,neg]}}$ & .483 \\ 
		Q$\mathtt{_{[bys,lie]}}$ & .400 & .706 \\ 
		Q$\mathtt{_{[foo,arg]}}$ &  & -.456 \\ 
		Q$\mathtt{_{[foo,pos]}}$ & .531 & -.412 \\ 
		Q$\mathtt{_{[foo,neg]}}$ & .643 \\ 
		Q$\mathtt{_{[foo,lie]}}$ & .359 & .589 \\ 
		Q$\mathtt{_{[cat,arg]}}$ & .481 & -.336 \\ 
		Q$\mathtt{_{[cat,pos]}}$ & .552 \\ 
		Q$\mathtt{_{[cat,neg]}}$ & .702 \\ 
		Q$\mathtt{_{[cat,lie]}}$ & .444 & .492 \\ 
		 \hline 		
	\end{tabular} 	}  	
	\end{center} 	 	
	\caption{Amount of loading for each variable on latent dimensions D1and D2.} 
\label{tab:CATPCA_loading} 
\end{table}   

The inspection of the loadings suggest the following characterization of the three latent dimensions.   
D1 receives the (substantial) contribution of all the variables, except Q$\mathtt{_{[foo,arg]}}$; see Table \ref{tab:CATPCA_loading}. It can therefore be interpreted as a sort of general ``persuasion acceptance'' dimension.   
D2 divides the variables into three groups; see Fig. \ref{img:comp_loading}. 

\begin{itemize} 
	\item	Variables with high positive loadings - namely, the lie strategy variables Q$\mathtt{_{[bys,lie]}}$, Q$\mathtt{_{[foo,lie]}}$ and Q$\mathtt{_{[cat,lie]}}$. 
	\item	Variables with high negative loadings (Q$\mathtt{_{[bys,arg]}}$, Q$\mathtt{_{[bys,pos]}}$, Q$\mathtt{_{[foo,arg]}}$, Q$\mathtt{_{[foo,pos]}}$, Q$\mathtt{_{[cat,arg]}}$) including all the cases of the argumentative strategy and two instances of the positive emotional one. 	\item	Variables with loadings close to zero (Q$\mathtt{_{[bys,neg]}}$, Q$\mathtt{_{[foo,neg]}}$, Q$\mathtt{_{[cat,pos]}}$ and Q$\mathtt{_{[cat,neg]}}$) consisting mainly of stimuli realizing the negative emotional strategy. 
\end{itemize} 

The opposition between the lie strategy and the argumentative one, along with the neutral role of the negative emotional strategy, suggest that D2 captures the effect that the truth-conditional value of what is said by the persuasive message has on subjects' responses. It can be of some interest, in this connection, that two of the three instances of the positive emotion strategy, Q$\mathtt{_{[bys,pos]}}$ and Q$\mathtt{_{[foo,pos]}}$, seem to part together with the argumentative, suggesting that subjects perceive/assign similar truth-conditional values to positive emotions in the bystander and in the footbridge scenarios; in the catastrophic one, though, the positive emotion strategy (Q$\mathtt{_{[cat,pos]}}$) becomes  ``truth-conditionally'' neutral and parts together with the negative emotion strategy.  
The other dimensions, not discussed here, account for progressively decreasing amounts of variance, scattering them on few variables. Finally, and importantly, the given picture is not affected by the Agent factor. As a consequence, it is not only the case that the nature of the persuasion agent fails to affect the ethical acceptability of the proposed persuasion situations; it also does not affect the structure of the attitude towards ethical acceptability itself. 

In conclusion, the analysis of the CATPCA results suggests the existence of a general `persuasion acceptance' latent dimension and of a second latent dimension, D2, accounting for the import of the `truth-conditional' value of the persuasion message. The poles of D2 are identified by the lie and by the argumentative strategy, respectively; the other strategies are either attracted towards one of the two poles, as it happens with the positive emotional strategy ($\mathtt{ST_{pos}}$) that parts with the classical one in $\mathtt{SC_{bys}}$ and $\mathtt{SC_{foo}}$, or simply do not contribute to D2, being truth-conditionally neutral - as is the case with the negative emotion strategy and the positive emotion strategy when used in the catastrophic scenario. It was not possible to shed more light on the role of the negative emotion strategy on the basis of the available data.   

\section{Discussion}

The cross-scenario differences in moral acceptability have similar direction as those reported in the literature for the direct action case, but different magnitudes. In particular, in 
the bystander scenario has a higher acceptability than in ours - 77\% in  the BBC survey\footnote{http://news.bbc.co.uk/2/hi/uk\_news/magazine/4954856.stm} and 90\% in \cite{mikhail2007universal} and \cite{hauser2006moral} - while the acceptability of the footbridge scenario sharply decreases when direct action is at stake - 10\% in \cite{mikhail2007universal,hauser2006moral} and 27\% in the BBC survey. 
The similar directions and the different magnitudes suggest a role for \emph{liability}: in traditional cases, the main character takes full responsibility for choosing between the alternative direct actions. 
In the persuasion case, in turn, the main character (the persuader) does not take a similar responsibility for the acts he/she/it intends the ``traditional'' actor to perform. Apparently, this lowers the overall acceptability of persuasive acts while reducing cross-scenario differences.  

The absence of differences due to the nature of the persuader (human or machine) can be interpreted as showing that judgments of moral acceptability address more persuasion acts than the actors performing them. This result is compatible with the suggested difficulty in identifying clear liabilities for the persuader: being not liable for what he/she/it says, the persuader retreats in the background and the persuasion act remains in the foreground. A different (but not necessarily alternative) explanation might appeal to the media equation framework \cite{Reeves1996}, with the  qualification that in this instance we would face the previously never considered case of machines assigned with identical moral obligations as humans. 

  An important finding of this paper is the decomposition of people's attitude towards persuasive messages into (at least) a general `attitude' component and a specific factor accounting for the truth-conditional import of the persuasion message. Importantly, the latter is not defined only with reference to the straightforward cases (the argumentative and the lie strategies) but it also includes the usage of positive emotions. 
Future work should aim at: replicating the present study to assess results' robustness; understanding better the role of negative emotions, which have somehow eluded our effort in the present work; widening the scope to include other persuasion strategies and dilemma scenarios. 
The import of our findings for computational work on persuasive system is manifold. In the first place, the overall low moral acceptability suggests care in the resort to persuasion by intelligent systems. The two latent dimensions underlying moral acceptability, in turn, suggest maximizing the impact of persuasion by targeting subjects who score high on them, calling the attention on a view of personalized persuasion whereby moral acceptability adds to sensitivity to persuasion. Finally, personalized persuasion could take advantage of studies addressing the dispositional nature (if any) of people's attitude towards moral acceptability by, e.g., addressing the personality traits (if any) underlying it and their relationships to the two latent dimensions we found. 

\section{Conclusions}
In this paper we have described experiments addressing ethical issues for persuasion systems and have discussed the results. Unfortunately, while sensitivities are great, not much experimental work is available on this topic. The little attention given so far to the theme has privileged the first of the two actions involved in persuasion - the action that the persuader intends the persuadee to make - over the communicative action that the persuader exploits for persuading.  For an intelligent, adaptive persuasive system instead, flexibility will mostly consist in adapting the persuasion strategy to the persuadee's characteristics and to the situation. 

We have followed a  behavioral approach, in the tradition of  so called natural ethics and moral dilemmas, to advance understanding 
users' moral acceptability of real systems' behavior. Moral dilemmas are useful because they stretch the situation and force a choice among otherwise ethically unacceptable outcomes. Our findings can be summarized as follows: (i) the overall acceptability of persuasion acts tends to be low.
(ii) Acceptability is affected by the type of strategy adopted, with those belonging to the validity domain scoring higher than emotional ones. (iii) People do not seem to be much concerned by persuading actor being a computer rather than a human. (iv) Validity seems to be one of the \emph{psychological} dimensions people use in their judgments, along with a general \emph{attitude towards persuasion} factor.
The results pave the way for a novel line of work contributing both to a deeper understanding of the ethical acceptability of persuasion acts, and to providing systems with the capability of choosing appropriate strategies for influencing people given the situation they are in and their personal dispositions. 

\bibliographystyle{aaai}
\bibliography{paper-grafia-etica}

\end{document}